\documentclass[aps,prb,twocolumn,showpacs,preprintnumbers,amsmath,amssymb,superscriptaddress]{revtex4}
\usepackage{graphicx}% Include figure files
\usepackage{dcolumn}% Align table columns on decimal point
\usepackage{bm}% bold math
\newcommand{\nc}{\newcommand}
\nc{\be}{\begin{equation}}
\nc{\ee}{\end{equation}}
\nc{\bea}{\begin{eqnarray}}
\nc{\eea}{\end{eqnarray}}
\nc{\bean}{\begin{eqnarray*}}
\nc{\eean}{\end{eqnarray*}}
\nc{\mb}{\mbox}
\nc{\rnc}{\renewcommand}
\nc{\vk}{\mb{\bf k}}
\nc{\vp}{\mb{\bf p}}
\nc{\vn}{\mb{\bf n}}
\nc{\vq}{\mb{\bf q}}
\nc{\rr}{\mb{\bf r}}
\nc{\vz}{\hat {\mb{\bf z}}}
\nc{\vj}{\mb{\boldmath$j$}}
\nc{\vg}{\mb{\boldmath$g$}}
\nc{\x}{\mb{\boldmath$x$}}
\nc{\A}{\mb{\boldmath$A$}}
\nc{\va}{\mb{\boldmath$a$}}
\nc{\vs}{\mb{\boldmath$\sigma$}}
\nc{\vpi}{\mb{\boldmath$\pi$}}
\nc{\nab}{\nabla}
\nc{\X}{\sf x}

\begin{document}

%\preprint{APS/123-QED}

\title{
Coupled charge and valley excitations in graphene quantum Hall ferromagnets
}

\author{Naokazu Shibata}
\affiliation{Department of Physics, Tohoku University, Aoba, Aoba-ku, Sendai
  980-8578 Japan} 
\author{Kentaro Nomura}
\affiliation{Department of Physics, Tohoku University, Aoba, Aoba-ku, Sendai
  980-8578 Japan} 

\date{21 May 2008}

\begin{abstract}
Graphene is a two-dimensional carbon material with a honeycomb lattice
and Dirac-type low-energy spectrum. In a strong magnetic field, where
Coulomb interactions dominate against disorder broadening, quantum Hall
ferromagnetic states realize at integer fillings. 
Extending the quantum Hall ferromagnetism to the fractional filling case of
massless Dirac fermions, we study the elementally charge excitations which
couple with the valley degrees of freedom (so-called valley skyrmions). With
the use of the density matrix renomalization group (DMRG) method, the
excitation gaps are calculated and extrapolated to the thermodynamic limit. 
These results exhibit numerical evidences and criterions of the skyrmion
excitations in graphene.
\end{abstract}

\maketitle

\noindent
\section{ Introduction}
%
% [IQHE in graphene]---------
A recent experimental realization of single-layer graphene sheets
\cite{Novoselov_2004} has made it possible to confirm a number
of theoretical predictions of intriguing electric properties of  
 massless Dirac fermion systems,\cite{Neto_2007}
including unconventional quantum Hall effects
(QHE)\cite{Novoselov_2005,Zhang_2005} 
with the half-integer Hall conductivity\cite{Dirac_qhe} 
\bea
\sigma_{xy} = \frac{4e^2}{h} (n +\frac{1}{2})
\eea
at $\nu=\pm2,\pm6,\pm10, \cdots$, where a factor $4$ is the Landau level (LL) 
degeneracy, accounting for spin and valley symmetry in graphene.
$\nu=2\pi\ell_B^2\rho$ is the filling factor, $\ell_B=\sqrt{\hbar/eB}$
is the magnetic length, $\rho$ is the carrier density measured from
charge neutral Dirac point. 

% [Additional QHE states]-----------
In addition to these unconventional quantized values, which can be solely 
understood on the basis of the massless Dirac fermion spectrum in a
magnetic field: 
\bea
E_n={\rm sgn}(n)\hbar v_F^{}\sqrt{2|n|}/\ell_B,
%+g\mu_B{\bf B}\cdot{\bf S},
\eea
recent experimental studies in a sufficiently strong magnetic field 
revealed new quantum Hall states at $\nu=0,\pm1,\pm4$,\cite{Zhang_2006} 
where the electron-electron interaction may play a crucial role. 
%
% [Energy scales in a B field]----------
Here relevant energy scales in graphene in a magnetic field are   
(i) Landau level (LL) separation between $n=0$ and $n=\pm 1$, 
$\hbar\omega_0^{} \equiv \sqrt{2}\hbar v_F^{}/\ell_B 
\simeq 400\sqrt{B[T]}[K]$, 
(ii) Zeeman coupling, $\Delta_z\equiv g\mu_B|{\bf B}|\simeq
1.5\times(B[T])[K]$, and  
(iii) the Coulomb energy, $e^2/\epsilon l_B\simeq 100\sqrt{B[T]}[K]$.
% [Experimental results]-------------- 
The activation energy measurements\cite{Zhang_2006} have shown that at
$\nu=\pm4$ the gap has linear $B$ dependence and reasonably corresponds 
to $\Delta_z$, indicating Zeeman spin splitting. 
At $\nu=\pm 1$, however, the gap is approximately scaled by $\sqrt{B}$, 
which indicates that the gap originates from the Coulomb interaction. 
%
% [Theory side]----------
%
There have been a number of theoretical investigations of these states, and 
there are two leading theoretical scenarios for the origin of the gap.
One is the quantum Hall ferromagnetism (QHF)\cite{QHF_review}
in which valley degrees of freedom, referred to as pseudospins,
spontaneously split via the exchange energy at {\it all integer} 
fillings.\cite{Nomura_2006,Alicea_2006,Goerbig_2006,Yang_2006,
Chakraborty_2007,Sheng_2007}   
Second is the spontaneous mass 
generation,\cite{Gusynin_2006,Herbut_2006,Fuchs_2006}
which predicts a gap {\it only} in the $n=0$ LL.
Although they are not entirely orthogonal to each other and 
the order parameters in the two theories 
can coexist,\cite{Yang_2007} excitation properties, including the existence of
gaps in $n\ne 0$ LLs, are different.  

In the QHF theory, the ground state at $\nu_n=1$, 
where $\nu_n$ is the filling factor for $n$th Landau level defined by
\bea
 { \nu}_n=\nu-4(n-1/2),
\eea
is fully spin and valley polarized, 
and the wave function can be represented by  
\bea
 |\Psi^{{ {\nu}_n}=1}_{\tau}\rangle
=\prod_{m}c^{\dag}_{m,\tau}|0\rangle, 
\eea 
where we assign the valley $K$ and $K'$ in graphene
as $z$-component of pseudospin $\tau=K$ or $K'$.
Excitations from the symmetry broken states are described by
(pseudo)spin wave and (pseudo)spin textures called 
skyrmions\cite{Sondhi_1993,Fertig_1994,Moon_1995,Rezayi_1991,Yoshioka_1998}
or other types, depending on the LL index $n$. 
Yang {\it et al.}\cite{Yang_2006} estimated skyrmion
excitation energy within the framework of Hartree-Fock (HF)
approximation. 

The fractional quantum Hall effects (FQHE) in graphene have been studied
by Apalkov and Chakraborty\cite{Apalkov_2006}. As a consequence of the
relativistic nature of electrons in graphene, the effective
electron-electron interactions in $n\ne 0$ LLs differ from that of
conventional two-dimensional
systems.\cite{Nomura_2006,Goerbig_2006,Apalkov_2006} 
With the use of the exact diagonalizations
they have calculated the magnetoroton energy at $\nu_n=1/3$ 
when spin and valley are fully polarized.
Their results show that the magnetoroton energy in the
$n=1$ LL is larger than that in the $n=0$ LL in finite systems. 
\cite{Apalkov_2006}
Nevertheless, the magnetoroton excitation is charge neutral and it is
not connected to the charged gap which is necessary to realize the FQHE. 
Moreover, the valley degrees of freedom which couple with charge
excitations must be taken into account 
since there is no external symmetry breaking effect for valley degrees 
of freedom. 
The excitation structures and the size of the 
activation gaps in the thermodynamic limit
have not yet been clarified systematically.\cite{Note_1} 

In this paper we extend the quantum Hall ferromagnetism to the fractional
filling case of graphene and show 
analogous properties of pseudospin ferromagnetism and
topological excitations in the fractional quantum Hall
effect (FQHE) states in graphene. 
We start with the projected Coulomb interaction Hamiltonian onto a
certain LL, and study charge and valley excitations,
where we treat the valley degrees of freedom  $K$ and $K'$ in 
the language of the pseudospin, while real spin degrees of freedom
are supposed to be frozen by the Zeeman splitting. 
We calculate the exact wave functions of QHF ground states and low-energy
excited states basing on the density matrix renormalization group (DMRG) 
method, and confirm the existence of the skyrmion excitations of 
valley degrees of freedom not only at integer filling $\nu=1$ but also at
fractional filling $\nu=1/3$ for the LL index $n=0$, 1 and 2.  
The skyrmion energies 
at these fractions are calculated and 
extrapolated to the thermodynamic limit. 
The existence of the finite gaps at $\nu_n=1$ in the $n\ne 0$ LLs 
strongly supports the QHF but not the mass generation scenarios. 
We note that the DMRG method is suitable for the present study, since we
need to treat a large number of basis of the many-body Hilbert space 
including the psudospin degrees of freedom. To the best
of our knowledge, there is no numerical work so far to estimate reliable
extrapolated energies of the skyrmion excitation in the thermodynamic
limit even in the conventional FQHE systems.\cite{Note_1} 

\begin{figure}[t]
\begin{center}
\includegraphics[width=0.45\textwidth]{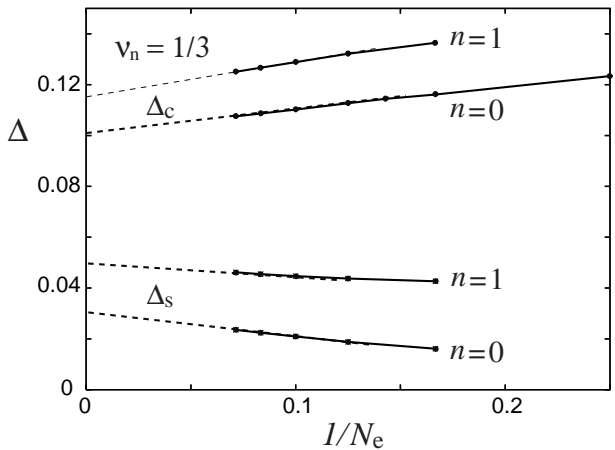}
\caption{
The pseudospin (valley) polarized excitation gap $\Delta_c$
and the pseudospin (valley) unpolarized (skyrmion)
excitation gap $\Delta_s$
at $\nu_n=1/3$ in the $n=0$ and 1 LLs.
}
\label{figure1}
\end{center}
\end{figure}

%\noindent
\section{model and Method}
We assume that the total magnetic field is strong enough to cause
complete spin splitting, and neglect spin flip excitations. We 
also neglect LL mixing for simplicity, although the ratio 
$(e^2/\epsilon l_B)/(\hbar\omega^{}_0)\simeq 0.25$ 
is not so small. The effect of LL mixing is discussed later. 
Taking account of valley degrees of freedom, 
we apply the DMRG method\cite{white,shibata1,shibata2} on the spherical
geometry.\cite{Haldane_1983,QHE_text,Fano_1986,Morf_2002,Feiguin_2007}

The projected Hamiltonian onto the $n$th Landau level is written
as\cite{Haldane_1983} 
\bea
 H^{(n)}=\sum_{i<j}\sum_{m}V^{(n)}_mP_{ij}[m],
\label{hamiltonian}
\eea
where $P_{ij}[m]$ projects onto states in which particles $i$ and $j$
have relative angular momentum $\hbar m$, and $V^{(n)}_m$ is their
interaction energy in the $n$th Landau level.\cite{Haldane_1983}
Using the {\it relativistic} form factor in the $n$th Landau
level\cite{Nomura_2006}
\bea 
 F_0^R(q)=L_0\left({q^2}/{2}\right)
\eea 
and 
\bea
 F_{n\neq 0}^R(q)=(1/2)\left[L_{|n|}\left({q^2}/{2}\right)
+L_{|n|-1}\left({q^2}/{2}\right)
\right],
\eea
the pseudopotentials\cite{QHE_text,Haldane_1983} are given by
\bea
V^{(n)}_m=\int_0^{\infty}\frac{dq}{2\pi}qV(q)e^{-q^2}[F^R_n(q)]^2L_m(q^2).
\eea
Here $L_n(x)$ are the Laguerre polynomials. The corresponding integrals
for electrons on the surface of a sphere which are used in the
present work are described in
Refs.~\onlinecite{Haldane_1983,Fano_1986,Morf_2002}.  
Note that our Hamiltonian Eq.(\ref{hamiltonian}) has SU(2) symmetry in
the valley degrees of freedom.  
A symmetry breaking correction to Eq.(\ref{hamiltonian}) which stems
from the honeycomb lattice structure of
graphene\cite{Alicea_2006,Goerbig_2006} is order of $a/\ell_B$ ($a$
being a lattice spacing) in units of $e^2/\epsilon \ell_B$ and neglected
in the following. 

We calculate the ground state wave function using the DMRG
method,\cite{white} which 
is a real space renormalization group method combined with the exact
diagonalization method. The DMRG method provides the low-energy
eigenvalues and corresponding eigenvectors of the Hamiltonian within a
restricted number of basis states. 
The accuracy of the results is systematically controlled by the
truncation error, which is smaller than $10^{-4}$ in the present
calculation. We investigate systems of various sizes with up to 40
electrons in the unit cell keeping 1400 basis in each
block.\cite{shibata1,shibata2}

\begin{figure}[t]
\begin{center}
\includegraphics[width=0.45\textwidth]{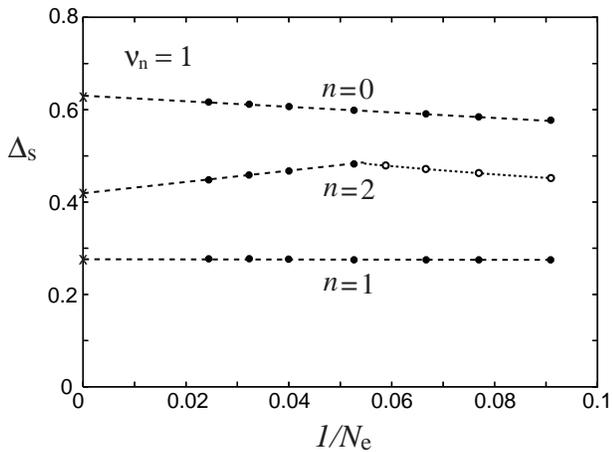}
\caption{
The pseudospin unpolarized (skyrmion) excitation gap 
$\Delta_s$ at $\nu_n=1$ in the $n=0$, 1 and 2 LLs.
The crosses on the vertical axis represent the results 
obtained by HF calculations.\cite{Yang_2006}
The pseudospin polarized excitation gaps in $n=2$ LL
are plotted by open circles. 
}
\label{figure1}
\end{center}
\end{figure}

%\noindent
%{\em Charged excitations}---
In the sphere geometry, the pseudospin (valley) polarized 
ground state at $\nu=1/q$ ($q$ being an odd integer), 
the Laughlin state,\cite{QHE_text} realizes 
when the total flux $N_{\phi}$ is given
by\cite{Haldane_1983} 
\bea
 N_{\phi}(\nu,N_e)=\nu^{-1}(N_e-1),
\eea
where $N_e$ is the number of electrons in the system.
Elementary charged excitations from this pseudospin polarized 
ground state correspond to the ground state configurations 
of the system with additional/missing flux $\pm 1$. 
In the following, we study two types of excitations:
Laughlin's quasiholes (quasiparticles)\cite{QHE_text} and 
skyrmion quasiholes (quasiparticles). 
\cite{Sondhi_1993,Fertig_1994,Moon_1995,Rezayi_1991,Yoshioka_1998}
Laughlin's quasiholes (quasiparticles) correspond to the
(pseudo)spin polarized excitations with $\pm 1$ flux,
whose creation energy is given by  
\bea
\Delta_c^{\pm}=E(N_{\phi}\pm 1,P=1)-E(N_{\phi},P=1),
\eea
 where $\pm$ represents quasiholes and quasiparticles, respectively,
 and $P$ is the polarization ratio of the pseudospin, i.e., $P\equiv
 (N_K-N_{K'})/(N_{K}+N_{K'})$ with $N_K$ $(N_{K'})$ 
being the number of electrons in $K$ $(K')$ valley.

Skyrmion quasiholes (quasiparticles) correspond to
the (pseudo)spin singlet excitations, and their 
creation energy is given by
\bea
\Delta_s^{\pm}=E(N_{\phi}\pm 1,P=0)-E(N_{\phi},P=1),
\eea
which could be smaller than $\Delta_c^{\pm}$.

The activation energy, referred to as the gap in the following, 
is given as a sum of these quasihole and quasiparticle energies,
$\Delta_c=\Delta_c^{+}+\Delta_c^{-}$ for pseudospin polarized
excitations, and $\Delta_s=\Delta_s^{+}+\Delta_s^{-}$ for 
pseudospin unpolarized excitations. 

%\noindent
\section{ Results}
Figure 1 shows $\Delta_c$ and $\Delta_s$ at $\nu_n=1/3$
as a function of $1/N_{e}$. 
In the $n=0$ LL, the pseudospin (valley) polarized excitation 
gap $\Delta_c$ is 0.101 $e^2/(\epsilon l_B)$ in the thermodynamic limit 
in good agreement with the previous work.\cite{Morf_2002}
In the $n=1$ LL, $\Delta_c$ is 0.115,
which is larger than 0.101 in the $n=0$ LL. 
This enhancement of $\Delta_c$ stems from the unique 
properties of the pseudopotentials in $n=1$ LL for 
relativistic particles; $V_1^{(1)}>V_1^{(0)}$.\cite{Apalkov_2006}
The magneto-roton excitation 
energy in the $n=1$ LL is also larger than that of the $n=0$ 
LL.\cite{Apalkov_2006}

The larger excitation gap in the higher LL has also 
been obtained for the pseudospin (valley) unpolarized 
excitations $\Delta_s$; the  unpolarized 
excitation gap $\Delta_s$ in the $n=1$ LL is 
0.05 $e^2/(\epsilon l_B)$, which is 
larger than that in the $n=0$ LL, $\Delta_s=0.03$. 
We note that similar analysis has been done 
in Ref.~\onlinecite{Toke_2006} with qualitatively different results.  

The results for $\nu_n=1$ are shown in Fig.2.
The extrapolated value of the 
pseudospin unpolarized excitation gap $\Delta_s$ for 
$n=0$ LL is 0.63 $e^2/(\epsilon l_B)$, which
is consistent with the previous HF studies and
exact diagonalization studies.\cite{Sondhi_1993} 
Even in the $n=1$ and $n=2$ LLs, $\Delta_s$ are 
almost the same with the HF results obtained in the 
four-component model for graphene.\cite{Yang_2006} 
These results show that the HF trial states for skyrmion 
excitation is essentially correct. 

In the above analysis, we have neglected the effect of LL mixing. 
When we study higher LLs, we need to take account of LL mixing, 
because LL spacing of graphene decreases with the increase 
in its index $n$.
In conventional two-dimensional systems, it has been argued that 
LL mixing reduces the activation gap in the fractional QHE 
regimes.\cite{Yoshioka_1986} 

\begin{figure}[t]
\begin{center}
\includegraphics[width=0.45\textwidth]{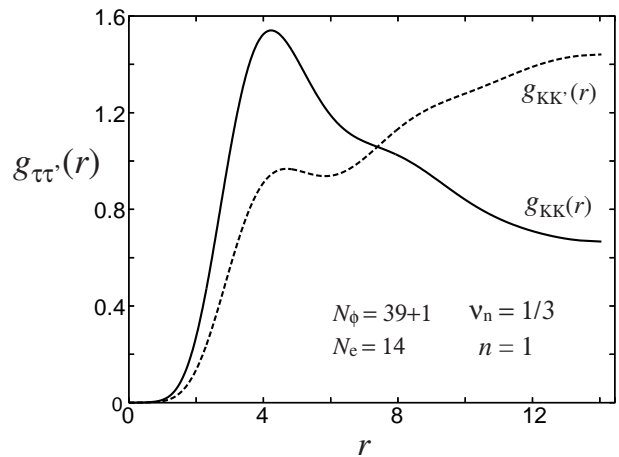}
\caption{
Two-particle correlation functions $g_{\tau\tau'}$ of quasihole 
skyrmion state at $\nu_1=1/3$. 
}
\label{figure1}
\end{center}
\end{figure}

\begin{figure}[t]
\begin{center}
\includegraphics[width=0.48\textwidth]{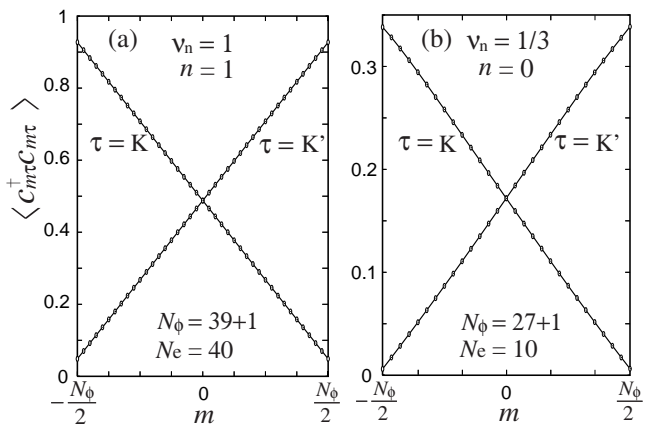}
\caption{
Expectation values, $\langle c_{m\tau}^{\dag}c_{m\tau}^{}\rangle$,
in pseudospin (valley) unpolarized excited states with one extra 
flux at $\nu_n=$1 and 1/3.  
}
\label{figure1}
\end{center}
\end{figure}

%\noindent
%{\em Two-particle correlation function}---
To study the skyrmion-like structure in the 
pseudospin unpolarized excited states, 
we have calculated the two-particle correlation
functions $g_{\tau\tau'}(r)$.\cite{Haldane_1983,QHE_text,Yoshioka_1998}
Figure 3 shows $g_{\tau\tau'}(r)$ for the pseudospin unpolarized 
quasihole state at $\nu_1=1/3$. 
We find clear peak structure around the origin $r=0$  
for the electrons in the same valley, $g_{KK}(r)$.
On the other hand, the maximum appears at the opposite 
side on the sphere($r=\pi R$) for the
electrons in different valleys, $g_{KK'}(r)$.
These structures are consistent with the 
skyrmion-like pseudospin structure.

\begin{figure}[!t]
\begin{center}
\includegraphics[width=0.48\textwidth]{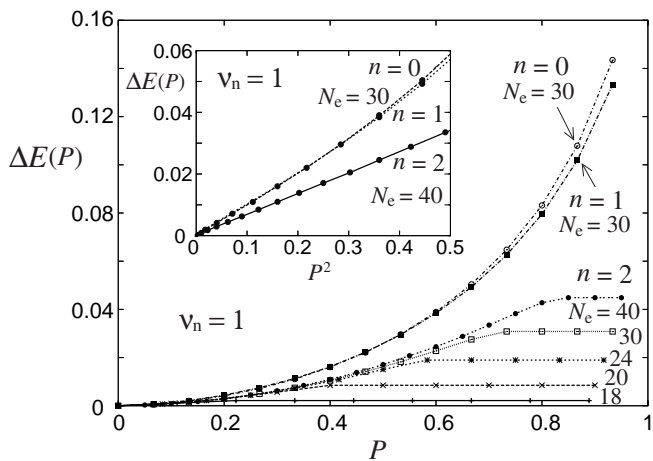}
\caption{
The polarization ratio $P$ dependence of the 
quasihole skyrmion energy
$\Delta E(P)=E(N_{\phi}+1,P)-E(N_{\phi}+1,P=0)$
at $\nu_n=1$ in the $n=0$, 1 and 2 LLs. 
}
\label{figure1}
\end{center}
\end{figure}

The HF trial state of quasihole skyrmions at $\nu_n=1$ is 
written in the form: 
\bea
 |\Psi_{sk}\rangle
=\prod_{m=-N_{\phi}/2}^{N_{\phi}/2}
[{\alpha}_m^{}
c_{m K}^{\dag}+{\beta}_m^{}c_{m+1 K'}^{\dag}]|0\rangle, 
\eea
where
$\langle c_{mK}^{\dag}c_{mK}^{\ }\rangle=|\alpha_m|^2$ and
$\langle c_{mK'}^{\dag}c_{mK'}^{\ }\rangle=|\beta_{m-1}|^2$.
To see the relation to this trial state,
we have calculated the expectation values 
$\langle c_{m\tau}^{\dag}c_{m\tau}^{\ }\rangle$
from the wave function obtained in the present DMRG study.
The results shown in Fig.4 indicate they are approximately given by 
$\langle c_{m\tau}^{\dag}c_{m\tau}^{\ }\rangle=1/2 \mp m/N_{\phi}$
for $\tau=K$ and $\tau=K'$, respectively.
Similar results are also obtained for $\nu_n=1/3$ in 
the $n=0$ and 1 LLs. 

The skyrmion excitations at $\nu_n=1$ are stable up to $n=2$ LLs, as 
shown in Fig.5, in which the polarization energies of the 
quasihole skyrmions $\Delta E(P)=E(N_{\phi}+1,P)-E(N_{\phi}+1,P=0)$
are plotted. 
The clear $P^2$-dependence shown in Fig.5 and its inset
confirms that the pseudospin singlet state is the lowest excitation 
for $n=0$, 1 and 2 LLs.
In the $n=2$ LL, the minimum at $P=0$ appears only for 
systems whose number of electrons is larger than 16; that means 
the skyrmion excitations are stable only for large systems. 
This is also shown in Fig.2 as an upward cusp in $\Delta_s$ 
at $N_e \sim 18$;
the pseudospin singlet state becomes the lowest when the number of 
electrons exceeds 18, while the pseudospin polarized states
marked by the open circles become the lowest for systems with 
$N_e < 18$. 

In higher LLs ($n \ge 3$), pseudospin singlet excitations
have not been seen as the lowest excitation within our 
calculations up to 40 electrons. 
These results are qualitatively consistent with the HF 
analysis in Ref.~\onlinecite{Yang_2006}, which predicts 
the skyrmion excitation gap is smaller than the 
HF quasiparticle gap only for 
lower LLs below $n=3$. 
The skyrmion excitation gap in $n=3$ LL is
close to the HF quasiparticle gap and it
is difficult to stabilize skyrmions in $n=3$ LL.

%\noindent
\section{ Discussion}
Our DMRG calculation confirms the valley polarized ground state 
at $\nu_n=1$ for $n \le 2$ LLs and at $\nu_n=1/3$ for $n \le 1$ LLs.
The (pseudo)spin polarized ground state at $\nu_n=1/q$ can be 
written simply as Laughlin's Jastrow function,\cite{QHF_review} and 
elementally charge excitations are obtained by
increasing or decreasing the flux quantum number $N_{\phi}$ by 1.  
We have studied both (a) pseudospin polarized excitations (Laughlin's
quasiholes and quasiparticles) and (b) pseudospin unpolarized excitations
(quasihole skyrmions and quasiparticle skyrmions). 
The activation energies obtained in finite systems are extrapolated to
the thermodynamic limit, which give theoretical predictions for future
experimental studies of the fractional quantum Hall states in graphene.  
In a high quality graphene sample, the $\nu_n=1$ QH states have been
observed in the $n=0$ LL.\cite{Zhang_2006} Very recently, 
suspended graphene sheets
with metallic contacts were succeeded.\cite{Bolotin_Du_2008}
The reported mobility $~ 2\times 10^{5}\ {\rm cm^2/Vs}$ is close to 
the crudely obtained critical value where the FQHE realizes within 
available magnetic field $\lesssim 50$ T, estimated from
the theory described in Ref.~\onlinecite{Nomura_2006} and the skyrmion
gap $\Delta_s \simeq 0.05 (e^2/\epsilon l_B)$ obtained in the present work.

\vspace{0.8mm}
%\noindent
%{\em Acknowledgments}--
\section*{Acknowledgment}
The present work was supported by Grant-in-Aid No. 18684012 from MEXT, Japan.

\end{document}